\documentclass{article}

\usepackage{graphicx}
\usepackage{float}
\usepackage[
singlelinecheck=false 
]{caption}

\begin{document}
\title{\bf On null geodesics and shadow of hairy black holes in Einstein-Maxwell-dilaton gravity}
\author{{Mohaddese Heydari-Fard$^{1}$ \thanks{Electronic address: m\_heydarifard@sbu.ac.ir}, Malihe Heydari-Fard$^{2}$\thanks{Electronic address: heydarifard@qom.ac.ir} and Hamid Reza Sepangi$^{1}$\thanks{Electronic address: hr-sepangi@sbu.ac.ir}}\\ {\small \emph{$^{1}$ Department of Physics, Shahid Beheshti University, Evin 19839, Tehran, Iran}}
\\{\small \emph{$^{2}$ Department of Physics, The University of Qom, 3716146611, Qom, Iran}}}

\maketitle

\begin{abstract}
The time-like and null-like geodesics around compact objects are some of the best tools to classify and understand the structure of a space-time. In this paper, we study the null geodesics around charged static dilaton black holes in Einstein-Maxwell-dilaton gravity. The physical parameters for non-radial geodesics including the effective potential, effective force, radius of the photon sphere and impact parameter are obtained and effects of the charge parameter and  dilaton coupling constant on these quantities are studied. Possible photon motions for different values of the impact parameter are analyzed and unstable circular orbits and unbounded orbits are plotted. These results are compared to that of the Schwarzschild, Reissner-Nordstrom and Gibbons-Maeda-Garfinkle-Horowitz-Strominger (GMGHS) black holes. Also, we study the shadow cast by a dilaton black hole and investigate how the dilaton coupling affects the size of the black hole shadow. Finally, as an application of null geodesics, we calculate the deflection of light and investigate the effects of the model parameters on the bending angle.
\vspace{5mm}
\vspace{1mm}
\end{abstract}

\section{Introduction}
Black holes are intriguing  objects in our universe. Ever since Einstein predicted their existence in general relativity (GR), the physics of black holes has attracted considerable attention. However, the theoretical results alone are not complete until backed up by observational data. After a century of theoretical research, remarkable successes have been achieved in the strong field regime around astrophysical black holes in the past few years. Besides the discovery of gravitational waves from the merger of a black hole binary by the Virgo and LIGO collaborations \cite{GW}, another important milestone is the direct observation of a supermassive black hole at the core of the M87* elliptical galaxy by the Event Horizon Telescope (EHT) collaboration \cite{Akiyama1}--\cite{Akiyama3} which released the first image of a black hole shadow. If the light rays get too close to a black hole, they get strongly deflected, or even move along circular orbits on the photon sphere. This strong deflection along with the fact that nothing can come out of the black hole, make the black hole seemed like a dark disk in the sky which is called the black hole shadow. The first study of black hole shadow for a Schwarzschild black hole and a rotating Kerr space-time was carried out in \cite{Synge} and \cite{Bardeen}, respectively. Such studies have received  significant attention in recent years and have been widely investigated in modified theories of gravity \cite{shadow1}--\cite{shadow15}.
Therefore, the study of null geodesics around black holes would help us to understand the properties of such objects and the geometric properties of the corresponding space-times. Also, such studies are useful for calculating  other related observable quantities such as gravitational lensing and deflection angle of light.

An effective way to understand the structure of a space-time is to study the geodesics that have exact analytical solutions. However, it is not always possible to solve the geodesic equations analytically so that numerical solutions would be in order. In such  cases, one can qualitatively analyse the behavior of geodesics  from the effective potential for the radial motion. The first exact solution of geodesic equations in the space-time of a Schwarzschild black hole was obtained in terms of elliptic functions by Hagihara in 1931 \cite{Hagihara}, followed by others who, over the years, have presented analytical solutions of the geodesic equations  of  such a metric \cite{Sch1}--\cite{Sch5}. Exact solutions of geodesic equations in the space-time of Schwarzschild, Kerr, Kerr-(anti) de Sitter, Reissner-Nordstrom, Reissner-Nordstrom-(anti) de Sitter and  Kerr-Newman have been also studied in \cite{Chandrasekhar}--\cite{KN}, respectively. In addition, the geodesic structure around black holes in modified theories of gravity have been extensively studied. For instance, the particle motion around black holes in $f(R)$ modified gravity and Horava-Lifshitz gravity have been studied in \cite{f(R)}--\cite{Horava4}. The analysis of null geodesics in brane world  scenarios and conformal Weyl gravity are  considered in \cite{brane1}--\cite{Weyl2}. In the space-time of Born-Infeld black holes, the null geodesics have been studied in \cite{Born1}--\cite{Born4}. The time-like and null-like geodesics in the background of quintessence black holes have been considered in \cite{qu1}--\cite{qu5}. The study of geodesic structure around hairy black holes has been carried out in \cite{hairy1}--\cite{hairy2}.  For the analysis of time-like and null-like geodesics in the background of wormhole geometries, see \cite{wh1}--\cite{wh3}.

It is generally accepted that the theory of GR is the most successful theory of gravity in offering a correct description of the universe, from the planetary motion to large scale structure. Despite these persuasive successes of GR, there are still many open problems including inflation, dark matter and dark energy. Also, at small scales where quantum gravity is important, there is no plausible theory of gravity. However, string theory, in a manner dictated by its low energy limit and by supplementing the usual Einstein-Hilbert action with higher-order curvature invariants, along with an extra scalar field non-minimally coupled to gravity, could open a way for quantum gravity \cite{Green}. In this regard, a well-known scenario in which a dilaton field is non-minimally coupled to the Maxwell field has been the focus of attention in the last decade and is known as the Einstein-Maxwell-dilaton (EMD) gravity. Static, spherically symmetric charged black hole solutions of the theory were initially found by Gibbons and Maeda \cite{Gibbons} and also independently by Garfinkle, Horowitz and Strominger \cite{Garfinkle} which has the following line element
\begin{equation}
ds^2=-\left(1-\frac{r_+}{r}\right)\left(1-\frac{r_{-}}{r}\right)^{\frac{1-\alpha^2}{1+\alpha^2}}dt^2+
\frac{dr^2}{\left(1-\frac{r_+}{r}\right)\left(1-\frac{r_{-}}{r}\right)^{\frac{1-\alpha^2}{1+\alpha^2}}}+
r^2\left(1-\frac{r_{-}}{r}\right)^{\frac{2\alpha^2}{1+\alpha^2}}\left(d\theta^2+\sin ^2\theta d\phi^2\right),
\label{00}
\end{equation}
where $\alpha$ is the dilaton coupling constant and $r_+$ and $r_-$ represent the radii of outer and inner horizons, respectively. In the special case of $\alpha=1$ (the simplified heterotic string), the metric corresponding to GMGHS black hole is given by
\begin{equation}
ds^2=-\left(1-\frac{2M}{r}\right)dt^2+\frac{dr^2}{\left(1-\frac{2M}{r}\right)}+r\left(r-\frac{Q^2}{M}\right)\left(d\theta^2+\sin ^2\theta d\phi^2\right),
\label{2}
\end{equation}
where $M$ and $Q$ are the ADM mass and electric charge of the black hole, respectively. The geodesic structure of massive and massless particles in the space-time of a GMGHS black hole has been extensively studied. For instance, the geodesic structure of test particles and light rays around a GMGHS black hole has been studied in \cite{Fernando}--\cite{n1}. Also, the null geodesics and motion of charged test particles around a magnetically charged GMGHS black hole have been discussed in \cite{GMGHS3} and \cite{GMGHS4}, respectively. The time-like and null-like geodesics around rotating dilaton black holes have been considered in \cite{rotating1}. The geodesic structure of normal and phantom EMD black holes presented in \cite{phantom} and EMD axion black holes in \cite{axion},  have also been investigated. The deflection of light and shadow of charged stringy black holes is studied in \cite{GMGHS5}. Alternatively, interesting physical aspects of dilaton black holes with metric given by equation (\ref{00}) have been widely studied.  The extension of this solution to slowly rotating dilaton black holes was carried out in \cite{1rotating}--\cite{3rotating}. The black hole superradiance, phase transition and quasinormal modes of dilaton black holes have been considered in \cite{superradiance}--\cite{qnm}. Shadow of EMd axion black holes and charged dilaton wormholes, thin accretion disks and black hole mergers in EMD gravity have been studied in \cite{shadow}--\cite{merger2}. Also, the light deflection of EM(anti)D black holes, using  Gauss-Bonnet theorem, was studied in \cite{deflection} together with the effect of dilaton parameter $\alpha$ on the bending of light. The motion of electric and dilatonic charged particles with arbitrary mass around dilaton black holes have also been studied \cite{Maki}. However, the motion of massless particles in the space-time of dilaton black holes with arbitrary values of $\alpha$ has not been attracting as much attention. Therefore, in this paper we consider charged static dilaton black holes and study  photon trajectories on the null geodesics and investigate the effects of both the dilaton coupling and charge parameter and compare the results to that of the Schwarzschild solution, Reissner-Nordstrom  ($\alpha=0$) and also GMGHS  ($\alpha=1$) black holes.

The structure of paper is as follows. In section 2, we briefly introduce the charged static dilaton black holes and some of their features. The radial geodesics and null geodesics with angular momentum are studied in detail in section 3. The study of the black hole shadow is done in section 4. In section 5 we study the bending of light in the space-time of dilaton black holes. Finally, conclusions are presented in section 6.

\section{Static charged dilaton black holes in EMD gravity}
The action of EMD gravity with an arbitrary dilaton coupling is given by
\begin{equation}
{\cal S}=\int d^4x \sqrt{-g}\left[R-2g^{\mu \nu}\nabla_{\mu}\Phi\nabla_{\nu}\Phi-e^{-2\alpha\Phi}F_{\mu \nu}F^{\mu \nu}\right],
\label{1}
\end{equation}
where $g$, $R$ and $\Phi$ are the metric determinant, scalar curvature and  dilaton field, respectively. Also, $F_{\mu\nu}=\partial_{\mu}A_{\nu}-\partial_{\nu}A_{\mu}$ is the strength of the Maxwell field with $A_{\mu}$ being the electromagnetic vector potential. The coupling constant $\alpha$ determines the strength with which the dilaton is coupled to the Maxwell field and without loss of generality we consider $\alpha$ to take positive values. The low energy limit of string theory corresponds to the case of $\alpha=1$ and $\alpha=\sqrt{3}$ represents the five-dimensional Kaluza-Klein theory. Moreover, in the case of $\alpha=0$ we obtain Einstein-Maxwell theory coupled to a scalar field where the static black hole solution is identical to the Reissner-Nordstrom solution of GR. Varying action (\ref{1}) with respect to metric, dilaton and Maxwell fields results in
\begin{equation}
G_{\mu\nu}=2\left[\nabla_{\mu}\Phi\nabla_{\nu}\Phi -\frac{1}{2}g_{\mu\nu}\nabla_{\rho}\Phi\nabla^{\rho}\Phi +e^{-2\alpha\Phi}\left(F_{\mu \rho}F^{\rho}_{\nu}-\frac{1}{4}g_{\mu\nu}F^2\right)\right],
\label{i1}
\end{equation}
\begin{equation}
\nabla_{\mu}\nabla^{\mu}\Phi = -\frac{\alpha}{2}e^{-2\alpha\Phi}F ^2,
\label{i2}
\end{equation}
\begin{equation}
\nabla_{\mu}\left(e^{-2\alpha\Phi}F ^{\mu\nu}\right)=0.
\label{i3}
\end{equation}
where we use the shorthand notation $F^2=F_{\rho\sigma}F^{\rho\sigma}$. As we mentioned before, the solution for a spherically symmetric static charged dilaton black hole with arbitrary values of $\alpha$ is given by \cite{Gibbons}--\cite{Garfinkle}
\begin{equation}
ds^2=-f(r)dt^2+\frac{dr^2}{f(r)}+R(r)\left(d\theta^2+\sin ^2\theta d\phi^2\right),
\label{3}
\end{equation}
where
\begin{equation}
f(r)=\left(1-\frac{r_+}{r}\right)\left(1-\frac{r_{-}}{r}\right)^{\frac{1-\alpha^2}{1+\alpha^2}},
\label{4}
\end{equation}
and
\begin{equation}
R(r)=r^2\left(1-\frac{r_{-}}{r}\right)^{\frac{2\alpha^2}{1+\alpha^2}}.
\label{5}
\end{equation}
The behaviour of the vector potential and the dilaton field is given by
\begin{equation}
A_{t} = \frac{Mv}{r},
\label{8}
\end{equation}
\begin{equation}
\Phi(r) = \frac{\alpha}{1+\alpha^2}\log{\left(1-\frac{r_{-}}{r}\right)}.
\label{9}
\end{equation}
Also the radii of outer and inner horizons are as follows
\begin{equation}
r_{+}= M[1+\sqrt{1-v^2(1-\alpha^2)}],
\label{6}
\end{equation}
\begin{equation}
r_{-}=\frac{ M(1+\alpha^2)[1-\sqrt{[1-v^2(1-\alpha^2)}]}{(1-\alpha^2)},
\label{7}
\end{equation}
where $M$ is the mass of the black hole and $v$ denotes the ratio of the electric charge to the black hole mass, $v=\frac{Q}{M}$. Clearly, from the reality of $r_+$ and $r_-$, constraint $1>v^2(1-\alpha^2)$ should be imposed. Also, for all values of $\alpha$, these solutions have an event horizon at $r=r_+$, but for any nonzero value of $\alpha$ the inner horizon at $r=r_-$ is a curvature singularity. Thus, these solutions describe black holes only when $r_-<r_+$ \cite{Garfinkle}. The extremal limit of the metric, for which $r_+=r_-$, is achieved for $Q_{\rm max}=M\sqrt{1+\alpha^2}$. Here, it should be mentioned that we will restrict our studies to the region $r_+<r<\infty$, because in this region the $t$ direction, $\partial/\partial t$, is time-like and $r$ direction, $\partial/\partial r$, is space-like, while in the region between the two horizons, $r_-<r<r_+$, $\partial/\partial t$ is space-like and $\partial/\partial r$ is time-like.

It is easy to see that in the limit $v=0$, with arbitrary values of $\alpha$, we have the Schwarzschild metric with event horizon at $r_+=2M$ and an intrinsic singularity at $r_-=0$. Also, in the case of $\alpha=0$ we obtain the Reissner-Nordstrom space-time metric with $r_{\pm}=M[1\pm\sqrt{1-v^2}]$. Although $r_-$ appears to be ill-defined at $\alpha=1$, it is well behaved in the limit of $\alpha\rightarrow 1$ and approaches a finite value. It is easy to see that in this case, $r_+=2M$, $r_-=\frac{Q^2}{M}$, the metric in equation (\ref{3}) reduces to GMGHS metric in equation (\ref{2}). Note that this solution is almost identical to the Schwarzschild black hole but the difference is that areas of spheres of constant $t$ and $r$ depend on $Q$. When $r=Q^2/M$ the area approaches zero and the surface is singular. For $Q^2\leq 2M^2$ the singular surface is inside the event horizon while for $Q^2=2M^2$ coincides with the horizon and a transition between the black hole and naked singularity occurs \cite{Garfinkle}.
The behaviour of the areal radius, $\sqrt{R(r_+)}$, as a function of the dilaton coupling $\alpha$ and charge parameter $v$ is shown in Figure 1.  It is seen that the areal radius decreases with increasing $v$ and increases very slowly with increasing $\alpha$. Also, it can be seen that in the Schwarzschild case with $v=0$ we have $\sqrt{R(r_+)}=2M$.

\begin{figure}[H]
\centering
\includegraphics[width=3.0in]{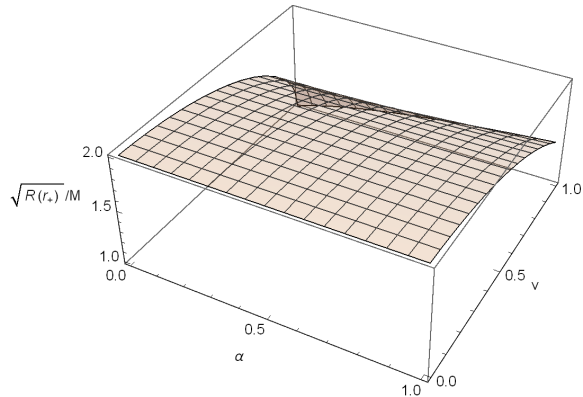}
\caption{\footnotesize The behaviour of the  areal radius at $r=r_+$ as a function of the charge parameter and dilaton coupling.}
\label{event-horizon}
\end{figure}

The dilaton charge is defined as  \cite{Gibbons}
\begin{equation}
D=-r^2 \Phi(r)_{,r}\mid_{r\rightarrow\infty}=-\frac{\alpha M [1-\sqrt{1-v^2(1-\alpha^2)}]}{(1-\alpha^2)}.
\label{10}
\end{equation}
We have plotted the behaviour of the dilaton charge as a function of dilaton coupling and charge parameter in Figure 2. As can be seen, with increasing both $v$ and $\alpha$ the dilaton charge increases. Moreover, for a Reissner-Nordstrom black hole with $\alpha=0$ and Schwarzschild solution ($v=0$) in GR, the dilaton charge is zero as one would expect.

\begin{figure}[H]
\centering
\includegraphics[width=3.0in]{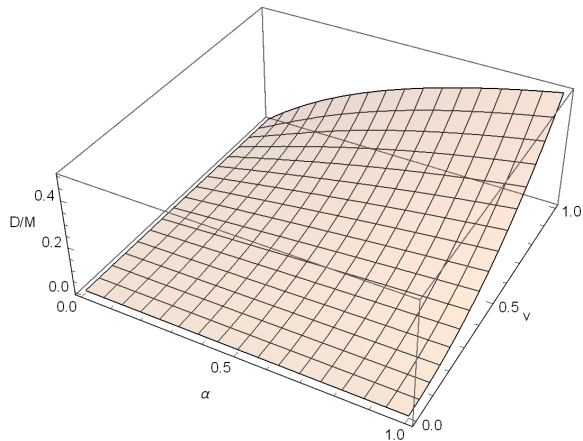}
\caption{\footnotesize The behaviour of the dilaton charge as a function of the charge parameter and  dilaton coupling.}
\label{dilaton-charge}
\end{figure}

\section{Null geodesics}
In this section, we aim to investigate the geodesic structure of massless particles in the space-time of dilaton black holes. Thus, we consider the Euler-Lagrange equation
\begin{equation}
\frac{d}{d\tau}\left(\frac{\partial \cal{L}}{\partial\dot{x^{\mu}}}\right)-\frac{\partial\cal{L}}{\partial x^{\mu}}=0,
\label{11}
\end{equation}
where $\tau$ is the affine parameter of the light rays and the Lagrangian for the metric in equation (\ref{3}) has the following form
\begin{equation}
{\cal{L}}=\frac{1}{2}g_{\mu\nu}\dot{x^{\mu}}\dot{x^{\nu}}=\frac{1}{2}\left(-f(r)\dot{t}^2+\frac{\dot{r}^2}{f(r)}+R(r)(\dot{\theta}^2+\sin^2\theta\dot{\phi}^2)\right)=0.
\label{12}
\end{equation}
Since the space-time is static and spherically symmetric,  we can restrict our study to the equatorial plane, $\theta=\frac{\pi}{2}$ without loss of generality.  Thus, by considering $\theta=\frac{\pi}{2}$ and $\dot{\theta}=0$, the Euler-Lagrange equations for $t$ and $\phi$ coordinates are given by
\begin{equation}
\dot{t}=\frac{E}{f(r)},
\label{13}
\end{equation}
\begin{equation}
\dot{\phi}=\frac{L}{R(r)},
\label{14}
\end{equation}
where $E$ and $L$, the energy and angular momentum of the photon, are the conserved quantities. Now, using the above equations, Lagrangian (\ref{12}) takes the form
\begin{equation}
E^2=\dot{r}^2+V_{\rm eff}(r),
\label{15}
\end{equation}
where the effective potential is given by
\begin{equation}
V_{\rm eff}=f(r)\frac{L^2}{R(r)}.
\label{16}
\end{equation}
In order to study the path of light rays, we need the relation between $r$ and $\phi$ which can be obtained by eliminating  $\tau$ from equations (\ref{14}) and (\ref{15})
\begin{equation}
\left(\frac{dr}{d\phi}\right)^2=\frac{R^2(r)}{b^2}-R(r)f(r),
\label{17}
\end{equation}
where we have substituted the effective potential from equation (\ref{16}) and defined the impact parameter, $b\equiv\frac{L}{E}$. In particular, for the light rays with critical value of the impact parameter, $b=b_c$, an unstable circular null trajectory occurs at the maxima of the effective potential $r=r_c$, known as the photon sphere \cite{sphere}. In what follows we investigate the radial null geodesics and null geodesics with angular momentum, separately.

\subsection{Radial null geodesics ($L=0$)}
For the radial motion with vanishing angular momentum, $L=0$, the effective potential is zero. So, from equations (\ref{13}) and (\ref{15}), one may find the differential equations governing the coordinate time $t$ and affine parameter $\tau$ as follows
\begin{equation}
\frac{dt}{dr}=\pm\frac{1}{f(r)},
\label{18}
\end{equation}
\begin{equation}
\frac{d\tau}{dr}=\pm\frac{1}{E},
\label{19}
\end{equation}
where the upper and lower signs denote the outgoing and ingoing motion, respectively. Assuming photons are at $r=r_i$ when $t=\tau=0$ and approaching $r=r_+$, we have plotted the behaviour of both $t$ and $\tau$ in the space-time of a charged dilaton black hole for ingoing photons in Figure 3. It is seen that in the affine parameter framework, the photons reach the horizon in a finite affine parameter while for the time coordinate it takes an infinite time, which is the same behavior as that in the Schwarzschild case \cite{Chandrasekhar}.

\begin{figure}[H]
\centering
\includegraphics[width=3.0in]{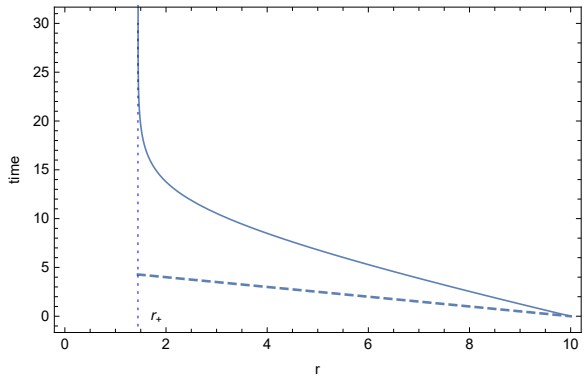}
\caption{\footnotesize The behaviour of the time coordinate $t$ (solid curve) and affine parameter $\tau$ (dashed curve) as a function of radius $r$. We have set $v=0.9$ and  $\alpha=0.1$. The vertical line represents the location of the event horizon at $r_+=1.44508$.}
\label{Time}
\end{figure}

\subsection{Geodesics with angular momentum ($L\neq0$)}
Now, we will consider the angular motion of photons in the space-time around EMD black holes.

\subsubsection*{A. The effective potential}
In this case, using equations (\ref{4})-(\ref{5}) and  (\ref{16}) the effective potential reads

\begin{equation}
V_{\rm eff}=f(r)\frac{L^2}{R(r)}=\left(1-\frac{r_+}{r}\right)\left(1-\frac{r_-}{r}\right)^{\frac{1-3\alpha^2}{1+\alpha^2}}\frac{L^2}{r^2}.
\label{20}
\end{equation}
For non-radial geodesics, it is convenient to set $L=1$ and thus from now on, we will consider $b=\frac{1}{E}$ in our calculations. We have plotted the effective potential in Figure 4. In the left panel, we have shown the effect of dilaton parameter on the potential. It is clear that by increasing $\alpha$, the maxima of effective potentials decrease so that a GMGHS black hole with $\alpha=1$ has the lowest maximum. For different values of $v$ with a fixed value of $\alpha$, the effective potential is plotted in the right panel. It is clear that with increasing $v$ the maxima of effective potentials assume higher values.
From equation (\ref{15}), one can see that the photon motion strongly depends on its energy levels. Thus, in order to discuss various cases of motions of photons, we have shown the effective potential and some energy levels in Figure 5. Assuming that light rays move in a radially inward direction, we summarize different kinds of motion as follows:

\begin{enumerate}
  \item  For $b<b_c$, the photon starts  at infinity and since it does not encounter the potential barrier,  falls into the black hole.
  \item The trajectories with impact parameter $b = b_c$ start falling inwards from $r > r_c$ and only ``reach'' the point $r = r_c$ in an infinite affine time, i.e. the
corresponding photons actually never reach this point but spiral with an infinite time around  the
photon sphere.
  \item In region 1, when $b>b_c$, the light rays that start from $r\geq r_0$, reach the potential barrier and thus are pushed back at $r=r_0$. However, when the photon starts at $r_+<r\leq r_1$ it will cross the horizon.
\end{enumerate}

\begin{figure}[H]
\centering
\includegraphics[width=3.0in]{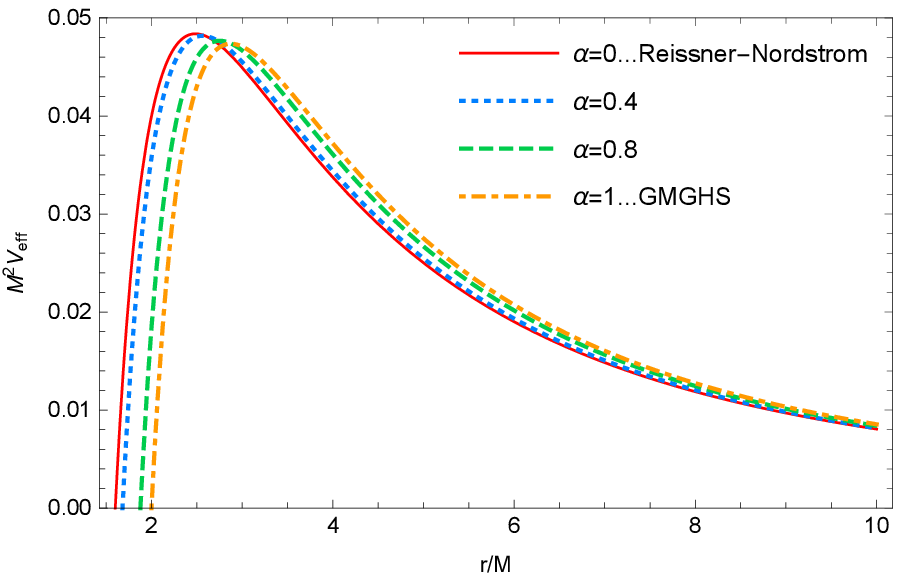}
\includegraphics[width=3.0in]{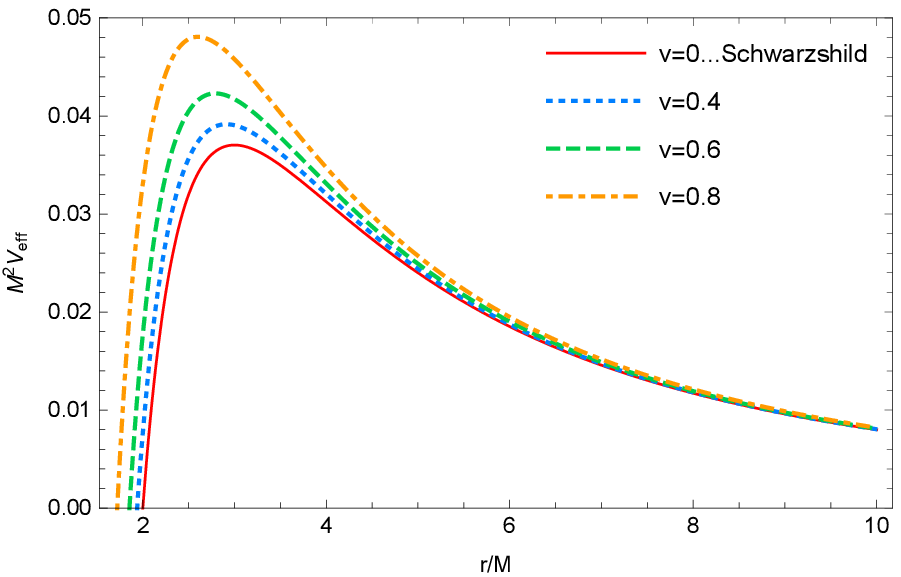}
\caption{\footnotesize The effective potential for massless particles around a static EMD black hole. In the left panel $V_{\rm eff}$ is shown for different values of $\alpha$ with $v=0.8$, and in the right panel for different values of $v$ for $\alpha=0.5$.}
\label{Effective-potential}
\end{figure}

\begin{figure}[H]
\centering
\includegraphics[width=3.0in]{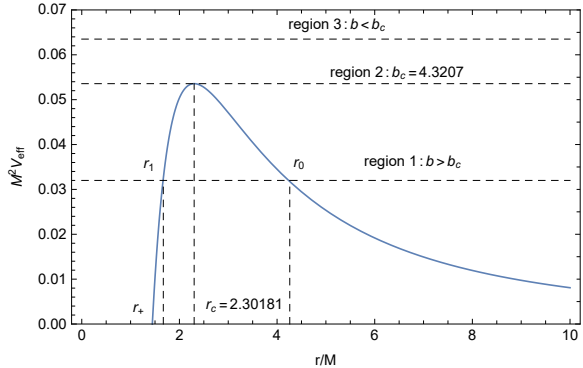}
\caption{\footnotesize The effective potential of photons around an EMD black hole and some energy levels. We set $\alpha=0.1$ and $v=0.9$. Region 2, corresponds to $V_{\rm eff}(r)=1/b_c^2$, while regions 1 and 3 correspond to $V_{\rm eff}(r)<1/b_c^2$ and $V_{\rm eff}(r)>1/b_c^2$ respectively.}
\label{potential-energy-levels}
\end{figure}

Another quantity of interest is the effective force on the photons which is given by
\begin{equation}
F(r)=-\frac{1}{2}\frac{dV_{\rm eff}}{dr}=\frac{\left(1-\frac{r_-}{r}\right)^{\frac{-4\alpha^2}{1+\alpha^2}}}{2r^5(1+\alpha^2)}\left[2(1+\alpha^2)r^2-3(1+\alpha^2)r_{+}r-(3-\alpha^2)r_{-}r+4r_+r_-\right].
\label{21}
\end{equation}
The factor $\frac{1}{2}$ appears because of the form of equation (\ref{15}) \cite{qu1}. In Figure 6, the total force on the photons as a function of $r$ is shown. As one expects from the shape of the potential, the effective force at the photon radius is zero. In other words, the point at which the force is zero, i.e. $r=r_c$, is important because it gives the location of the stationary point corresponding to circular geodesics. For $r_+<r<r_{c}$ the force is negative and photons feel an attractive force and as a result are pulled back towards the black hole and fall into it. It is interesting to note that for $r<r_{c}$ the force is attractive and that is due to the fact that we are studying the configuration between $r_+$ and $r_c$. However, for $r_c<r<\infty$, the force becomes positive, driving photons away from the black hole which is corresponding to photons being deflected at the turning point $r = r_0$. When $r\rightarrow\infty$ the force tends to zero and photons experience no force.
Also, the left panel of the figure shows the effect of the dilaton parameter on the effective force. As can be seen, by increasing $\alpha$ the effective force decreases. So, in the presence of dilaton hair, the force on the photon decreases so that a Reissner-Nordstrom black hole with $\alpha=0$, (the scalar-free solution), has the maximum value of the force. The effect of the charge parameter on the effective force is shown in the right panel. We see that by increasing $v$ the effective force also increases.
Meanwhile it is easy to see that in the limit of $\alpha\rightarrow 1$, equation (\ref{21}) reduces to the following relation
\begin{equation}
F_{GMGHS}(r)=\frac{1}{r^3(1-\frac{Q^2}{Mr})^2}\left[1-\frac{3M}{r}-\frac{Q^2}{2Mr}+\frac{2Q^2}{r^2}\right],
\label{new22}
\end{equation}
which is the effective force in the space-time of a GMGHS black hole. For $Q=0$ the above equation reduces to the effective force in the Schwarzschild space-time, that is
\begin{equation}
F_{Sch}(r)=\frac{1}{r^3}-\frac{3M}{r^4},
\label{22}
\end{equation}
where the first term represents the centrifugal force and the second term corresponds to the relativistic correction due to GR \cite{qu4}.

\begin{figure}[H]
\centering
\includegraphics[width=3.0in]{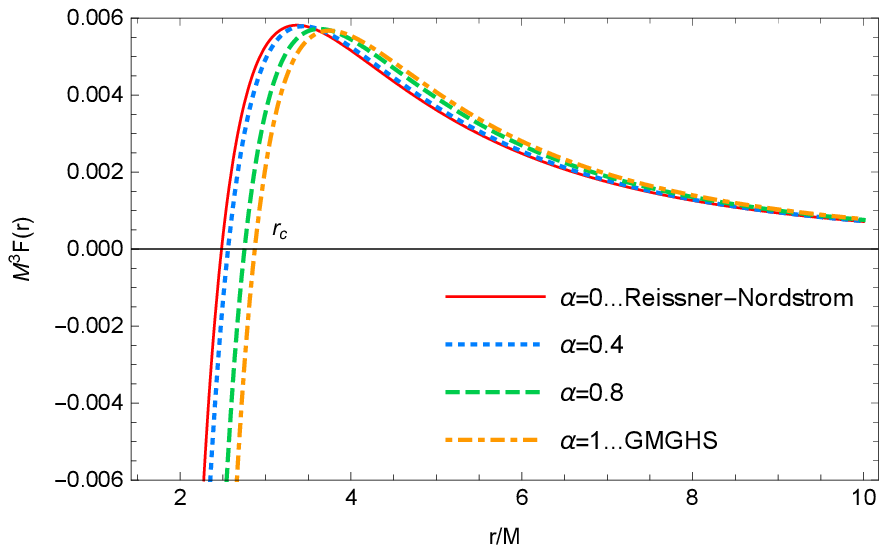}
\includegraphics[width=3.0in]{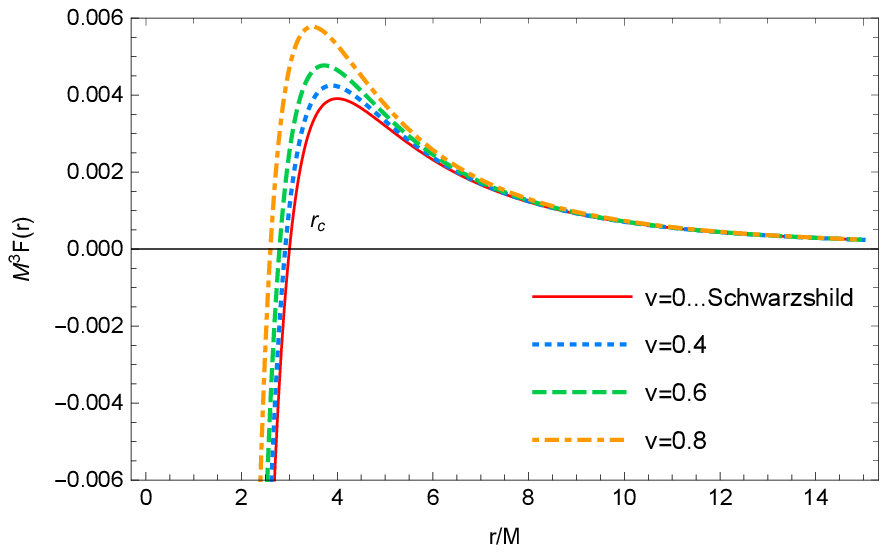}
\caption{\footnotesize The effective force as a function of $r$ for different values of $\alpha$ with $v=0.8$ (left panel), and for different values of $v$ with $\alpha=0.5$ (right panel).}
\label{Force}
\end{figure}

\subsubsection*{B. Circular orbits}
As was mentioned in the previous section, the circular orbit occurs at $r=r_c$ with $b=b_c$, so that the conditions for such a critical motion are given by
\begin{equation}
V_{\rm eff}=\frac{1}{b_c^2},
\label{25}
\end{equation}
\begin{equation}
\frac{dV_{\rm eff}}{dr}=0,
\label{26}
\end{equation}
which leads to the following relation
\begin{equation}
R(r)f'(r)-f(r)R'(r)=0.
\label{27}
\end{equation}
The above equation has two solutions for circular orbit radii as follows
\begin{eqnarray}
r_{c_{\pm}}&=&\frac{1}{2(1-\alpha^2)}\left[3M-2M\alpha^2-M\alpha^2\sqrt{1-v^2(1-\alpha^2)}\right.\nonumber\\
&\pm&\left.\sqrt{\left(3M-2M\alpha^2-M\alpha^2\sqrt{1-v^2(1-\alpha^2)}\right)^2-8v^2(1-\alpha^2)^2}\right].
\label{28}
\end{eqnarray}
We note that in the limit of $\alpha\rightarrow 1$ the above solution reduces to the radius of the circular orbit for the GMGHS black hole, namely equation (49) in \cite{GMGHS1}. In Figure 7, we have displayed both $r_{c_+}$ and $r_{c_{-}}$  together with the event horizon radius as a function of $v$ for $\alpha=0.9$. We see that for $v=0$, $r_{c_-}=0$ and $r_{c_+}=3M$ which is the radius of the unstable circular orbit for the Schwarzschild black hole. Moreover, it is clear that for all values of $v$ in this figure, $r_{c_{+}}$ is always larger than the event horizon radius,  $r_{c_{+}}>r_+$, while $r_{c_{-}}$ is smaller, $r_{c_{-}}<r_+$. Since, we will consider the photon motion in the region $r_+<r<\infty$, the radius denoted by $r_{c_{+}}$ is that of the unstable circular orbit and will be represented by $r_c$ from now on.
The dependence of the areal radius at $r=r_c$ on the dilaton coupling and charge parameter is shown in Figure 8. As can be seen, the areal radius is a decreasing function of $v$ and a very slowly increasing function of $\alpha$. Also, as is clear for an extremal Reissner-Nordstrom black hole with $\alpha=0$ and $v=1$, we have $r_c=2M$.

\begin{figure}[H]
\centering
\includegraphics[width=3in]{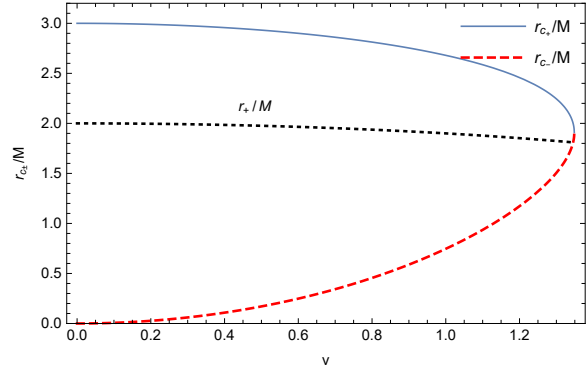}
\caption{\footnotesize The behaviour of $r_{c_{+}}$, $r_{c_{-}}$ and the event horizon radius $r_+$, as a function of $v$. The dilaton parameter is set to $\alpha=0.9$.}
\label{6rc}
\end{figure}

\begin{figure}[H]
\centering
\includegraphics[width=3in]{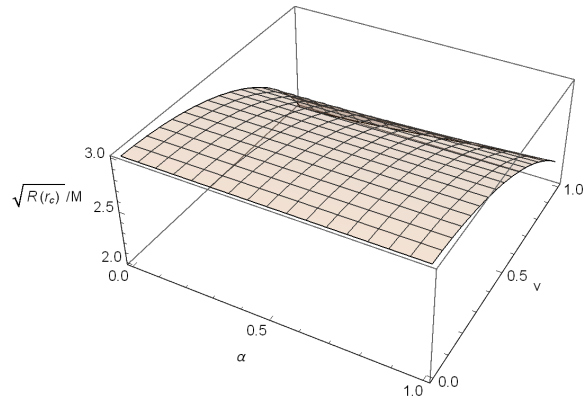}
\caption{\footnotesize The dependence of the areal radius at $r=r_c$  on the charge parameter and dilaton coupling.}
\label{photon-radius}
\end{figure}

Now, using equations (\ref{15})-(\ref{16}), the impact parameter of these unstable circular orbits is given by
\begin{equation}
b_c=\frac{L_c}{E_c}=\sqrt{\frac{R(r_c)}{f(r_c)}}=\frac{r_c}{\sqrt{(1-\frac{r_+}{r_c})(1-\frac{r_-}{r_c})^{\frac{1-3\alpha^2}{1+\alpha^2}}}}.
\label{29}
\end{equation}
We have presented the results of the event horizon radius, photon radius and  impact parameter of the photon sphere for different values of $v$ and $\alpha$ in Table 1 and 2. We see that for a fixed value of $\alpha$, as one increases the charge parameter $v$, the size of the event  horizon radius, photon radius and impact parameter $b_c$ decreases. However, Table 2 shows that for a given value of $v$,  increasing the dilaton parameter $\alpha$ causes the event horizon radius, photon radius and also value of the impact parameter to increases.

\begin{table}[H]
\centering
\caption{\footnotesize  The values of photon radius, $r_c$, impact parameter, $b_c$, and  the event horizon radius for different values of $v$, with $\alpha=0.2$. The first column with $v=0$ corresponds to the Schwarzschild black hole.}
\begin{tabular}{l l l l l l l l}
\hline
$$&$v=0$&$v=0.1$& $v=0.3$&$v=0.5$&$v=0.7$&$v=0.9$\\ [0.5ex]
\hline

$r_c/M$  &3       &2.9935  &2.9406 &2.8286 &2.6402 &2.3254\\
$b_c/M$  &5.1962  &5.1875  &5.1168 &4.9681 &4.7219 &4.3250\\
$r_+/M$  &2       &1.9952  &1.9558 &1.8718 &1.7277 &1.4716\\
$r_-/M$  &0       &0.0052  &0.0479 &0.1389 &0.2949 &0.5724\\

\hline
\end{tabular}
\end{table}

\begin{table}[H]
\centering
\caption{\footnotesize  The values of photon radius, $r_c$, impact parameter, $b_c$, and  the event horizon radius for different values of $\alpha$ with $v=0.6$. The results in the first and sixth columns correspond to that of Reissner-Nordstrom and GMGHS black holes, respectively.}
\begin{tabular}{l l l l l l l l l l l}
\hline
$$&$\alpha=0$&$\alpha=0.1$& $\alpha=0.3$&$\alpha=0.5$&$\alpha=0.7$&$\alpha=1$&$\alpha=1.3$&$\alpha=1.7$&$\alpha=2$\\ [0.5ex]
\hline

$r_c/M$  &2.7369     &2.7391   &2.7566   &2.7905  &2.8391  &2.9343 &3.0514 &3.2302 &3.3764\\
$b_c/M$  &4.8587     &4.8588   &4.8599   &4.8619  &4.8649  &4.8707 &4.8775 &4.8877 &4.8957\\
$r_+/M$  &1.8000     &1.8023   &1.8200   &1.8544  &1.9036  &2 &2.1173 &2.2963 &2.4422\\
$r_-/M$  &0.2000     &0.2017   &0.2156   &0.2427  &0.2818  &- &0.4547 &0.6098 &0.7370\\

\hline
\end{tabular}
\end{table}

\subsection{Analysis of geodesics in terms of the variable $u=\frac{1}{r}$}
The aim of this section is to study the geometry of null geodesics in the EMD black holes space-time. In order to do so, it is  more convenient to use the variable $u=\frac{1}{r}$. Therefore, we rewrite  equation (\ref{17}) in terms of  variable $u$
\begin{equation}
\frac{du}{d\phi}=\sqrt{\frac{(1-r_-u)^{\frac{4\alpha^2}{1+\alpha^2}}}{b^2}-(1-r_+u)(1-r_-u)u^2}\equiv\Phi(u).
\label{30}
\end{equation}
As can be seen, for $v=0$ the above equation reduces to $(\frac{du}{d\phi})^2=\frac{1}{b^2}+2Mu^3-u^2$ which is the equation of motion of photons in the space-time of the Schwarzschild black hole and, to Reissner-Nordstrom black hole for $\alpha=0$, for which it takes the form $(\frac{du}{d\phi})^2=\frac{1}{b^2}-Q^2u^4+2Mu^3-u^2$ \cite{Chandrasekhar}. Also, in the case of $\alpha=1$ with metric functions given by equation (\ref{2}), we find the corresponding result for GMGHS black holes, namely equation (41) in \cite{GMGHS1}.

As explained in the previous section, the photon motion depends on its energy level, see Figure 4. When $b=b_c$, the photons have an unstable circular orbit with radius $r=r_c$ and circle the black hole on the photon sphere. However, when $b<b_c$, the light rays will continue moving inward until captured by the black hole. Photons with $b>b_c$, will deflect at the point $u_0=\frac{1}{r_0}$ which is the root of the function $\Phi(u)$. In other words, the photon with $b>b_c$ reaches the turning point $u=u_0$ (or $r=r_0$) and is reflected, following an unbounded orbit.
Since equation (\ref{30}) does not have an analytical solution, we have numerically  plotted the resulting photon paths in Figure 9. The red curve shows photons coming from infinity and reaching the critical distance $r=r_c$, after which revolving around the black hole on the photon sphere. However, the blue curves show the light deflection around an EMD black hole for light rays with $b>b_c$. The null geodesics falling into the black hole is shown by the green curves.

\begin{figure}[H]
\centering
\includegraphics[width=2.50in]{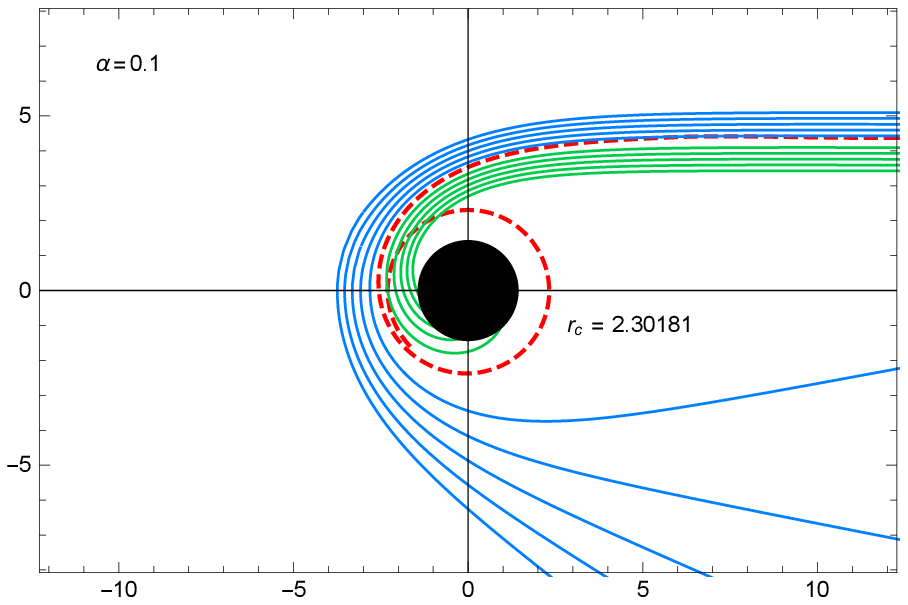}
\includegraphics[width=2.50in]{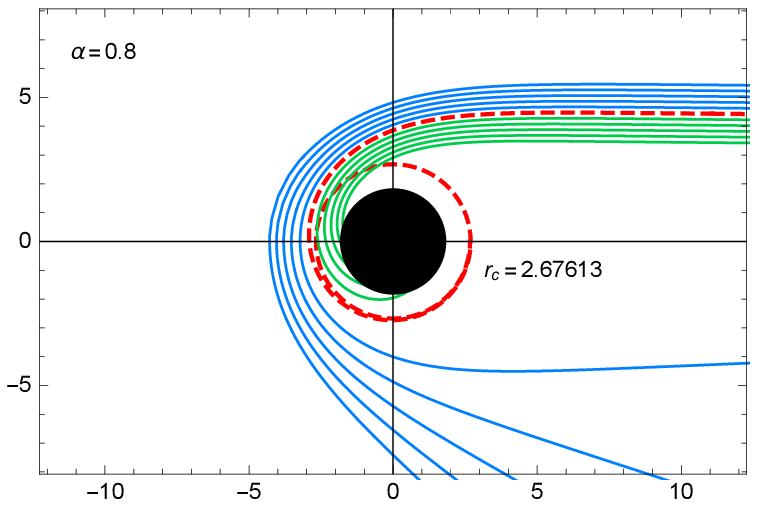}
\caption{\footnotesize The polar plot of null geodesics around an EMD black hole for $\alpha=0.1$ and $\alpha=0.8$, with $v=0.9$. The red, blue and green curves represent the null trajectories with $b=b_c$, $b>b_c$ and $b<b_c$, respectively. In each panel a black disk shows the event horizon at $r_+=1.44508$ and $r_+=1.84167$, respectively.}
\label{polar-plot}
\end{figure}

\section{Shadow of EMD black holes}
Let us now  consider the shadow cast by dilaton black holes. To this end, it is appropriate to introduce the celestial coordinates $X$ and $Y$ as follows \cite{Bardeen}
\begin{equation}
X=\lim_{r_*\rightarrow\infty}\left(-r_*^2\sin\theta_o\frac{d\phi}{dr}\right),
\label{n1}
\end{equation}
\begin{equation}
Y=\lim_{r_*\rightarrow\infty}\left(r_*^2\frac{d\theta}{dr}\right),
\label{n2}
\end{equation}
where $r_*$ is the observer distance to the black hole and $\theta_o$ is the angular coordinate of the observer, called the inclination angle. We restrict our study to the equatorial plane $\theta_o=\frac{\pi}{2}$, so that the radius of the shadow is given by
\begin{equation}
R_s=\sqrt{X^2+Y^2}=b_c.
\label{n3}
\end{equation}
In fact, it is found that in the particular case with $\theta_o=\frac{\pi}{2}$, for an asymptotically flat space-time with a line element in the form (\ref{3}), the radius of the black hole shadow is equal to the critical impact parameter. Also, it can be seen that for the Schwarzschild black hole, the radius of the shadow is $R_s=3\sqrt{3}\approx 5.1962$.

In Figure 10, we have shown the shadow boundaries of dilaton black holes for different values of $v$ and for changing  values of the dilaton parameter $\alpha$. We see that the shape of the black hole shadow is a perfect circle and for a fixed value of $\alpha$ the shadow size decreases by increasing $v$. The dilaton parameter runs from $\alpha=0$, corresponds to a Reissner-Nordstrom solution, to $\alpha=1$ for the case of a GMGHS black hole. The effect of charge and dilaton parameter on the shadow radius of a dilaton black hole is shown in Figure 11. It is seen that for a given value of $v$, the presence of the dilaton coupling $\alpha$ leads to  larger values for $R_s$ compared to the case of Reissner-Nordstrom black hole ($\alpha=0$) in GR. We note that the rate of increase in the shadow radius is larger for larger values of $v$. Also, for an extremal Reissner-Nordstrom black hole with $v=1$, we have $R_s=4M$. It is worth  mentioning that the measurements of the shadow size around the black hole may help to estimate the black hole parameters and probe the geometry of the background metric. However, for EMD black holes we are studying here, as the results in Tables 1 and 2 show, the size of the shadow, that is $R_s=b_c$, clearly has a strong dependency on the charge parameter $v$ while the dependency on $\alpha$ seems to be quite weak. In fact, for fixed values of $v$ and values of $\alpha$ from 0 to 2, the shadow size changes only by about 0.037. Therefore we find the effect of the dilaton parameter on the shadow size to be negligible in observational measurements.

On the other hand, it is expected that the astrophysical black holes have negligible electric charge so that the charge parameter $v$ is zero or small. Therefore, it may be difficult to constrain the dilaton parameter $\alpha$ from observations of the shadow size or gravitational lensing.

\begin{figure}[H]
\centering
\includegraphics[width=2.25in]{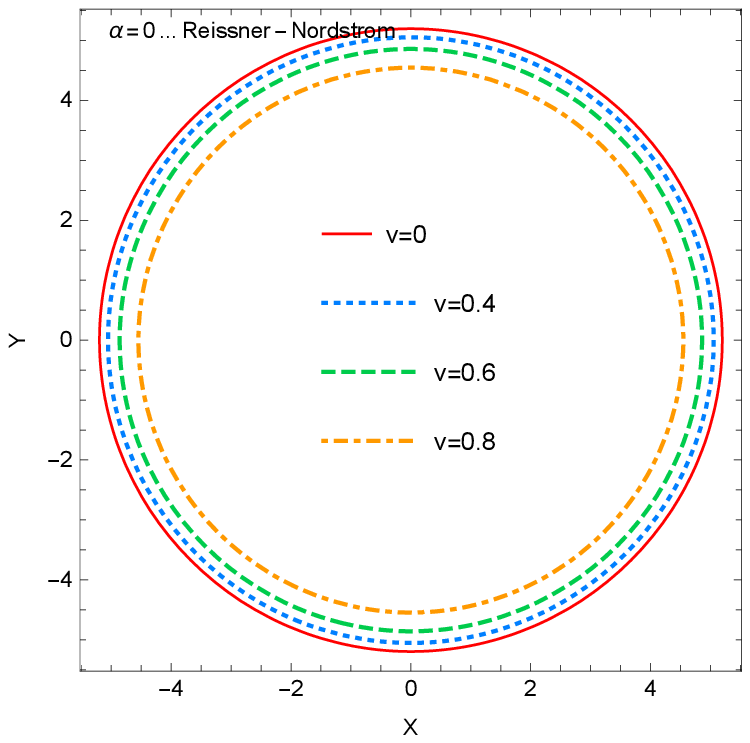}
\includegraphics[width=2.25in]{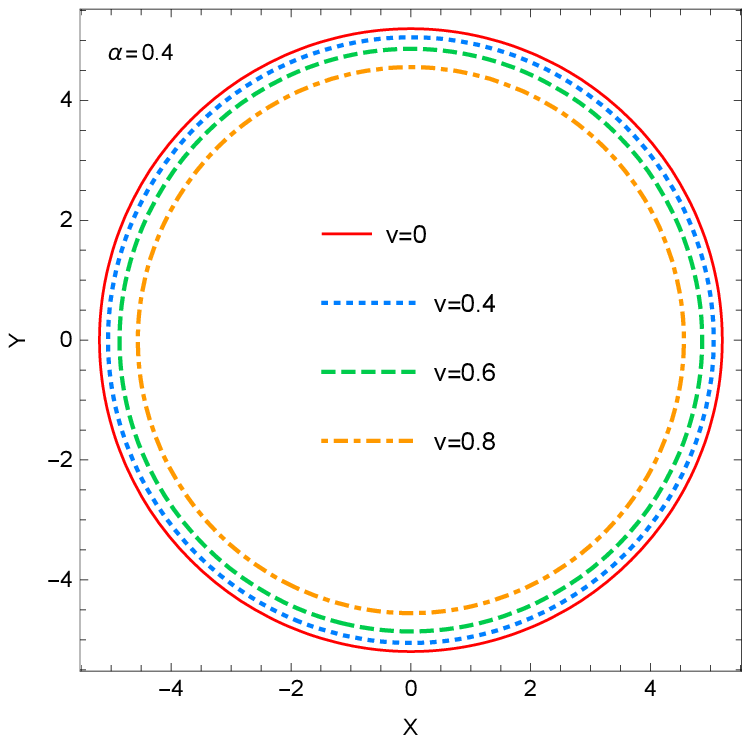}
\includegraphics[width=2.25in]{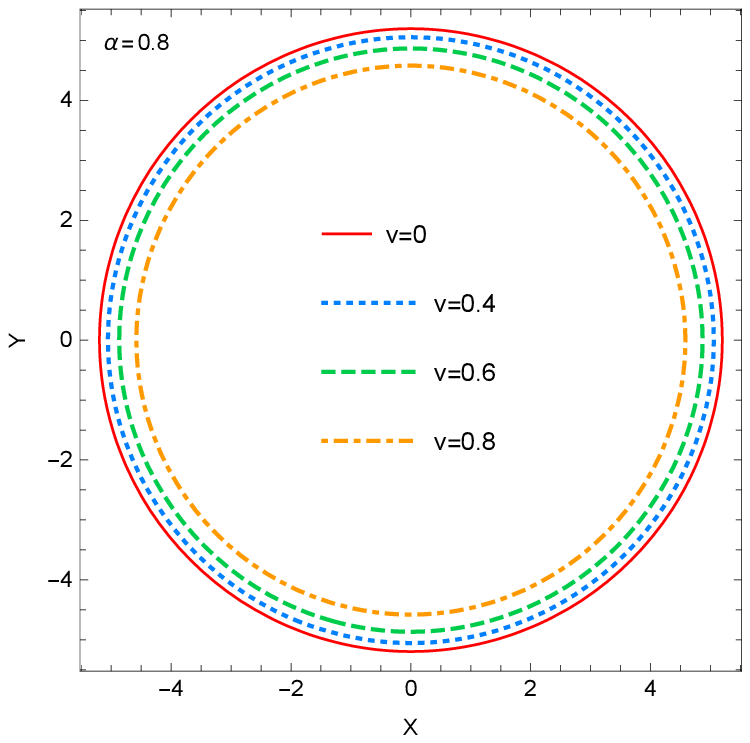}
\includegraphics[width=2.25in]{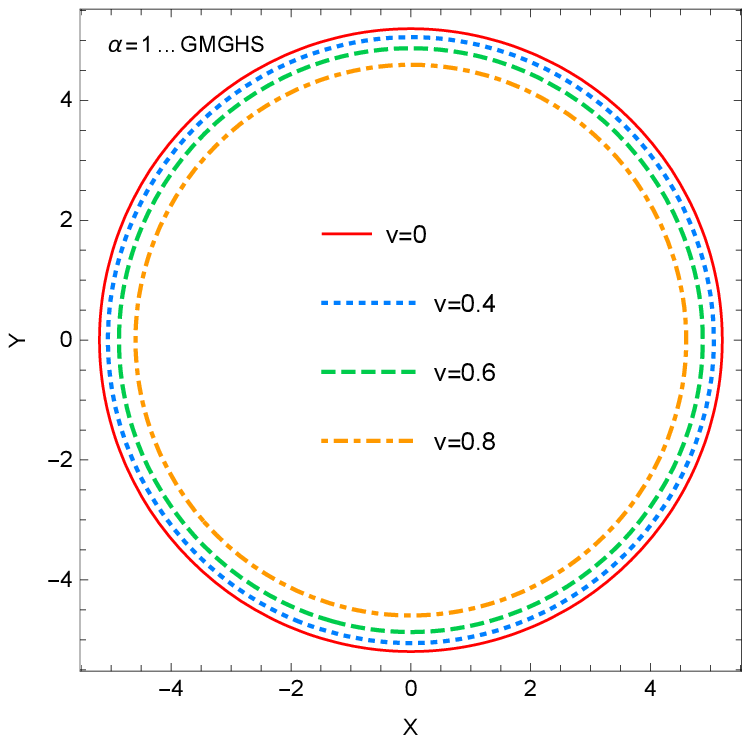}
\caption{\footnotesize The boundary of the shadow of a charged dilaton black hole for different values of the charge parameter $v$ and  dilaton coupling $\alpha$. In each panel the solid red curve shows the shadow of the Schwarzschild black hole. The unit of length along coordinate axes $\alpha$ and $\beta$ is $M$.}
\label{ss}
\end{figure}

\begin{figure}[H]
\centering
\includegraphics[width=2.83in]{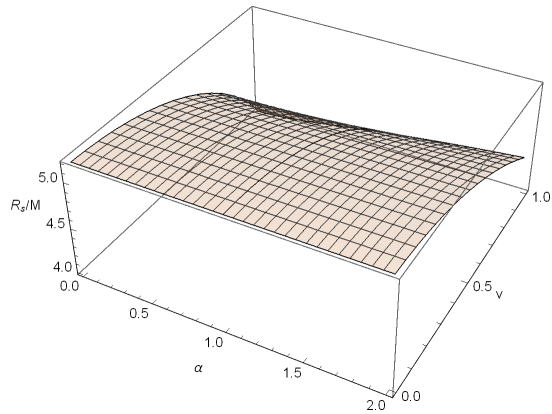}
\caption{\footnotesize The behaviour of the shadow radius as a function of charge parameter and of the dilaton coupling.}
\label{Rs}
\end{figure}

\section{Gravitational lensing by EMD black holes}
The study of light paths is important to investigate the gravitational lensing effects of compact objects. The problem of light bending in the space-time of a GMGHS black  and also EM(anti)D black hole using Gauss-Bonnet theorem has been considered in \cite{deflection}. Now, we focus attention on the gravitational lensing of a dilaton black hole and investigate the dependence of the bending angle on the dilaton coupling $\alpha$ and  charge parameter $v$. As was explained in section 3.3, when light rays with $b>b_c$  from infinity  encounter  the potential barrier, they are deflected at the turning point $r=r_0$, see Figure 4. Therefore, we first need  to find the closest distance $r_0$, which is the value of $r$ when $\frac{dr}{d\phi}=0$. From equation (\ref{17}) we have
\begin{equation}
\left(\frac{dr}{d\phi}\right)^2=\frac{r^4(1-\frac{r_+}{r})^{\frac{4\alpha^2}{1+\alpha^2}}}{b^2}-(r-r_+)(r-r_-)\equiv\Psi(r).
\label{31}
\end{equation}
In Figure 12, we have displayed  $\Psi(r)$ for different values of the impact parameter $b$. As one can see, there are two values of $r$ for which $\Psi(r)=0$. Since from Figure 4 the value of the turning point at $r=r_0$ is larger than the photon radius at $r=r_c$, we consider the larger roots in Figure 12 as the closest distance and identify them with $r_0$. Also, we note that with an increasing impact parameter the closest distance increases as well.

Using equation (\ref{30}), one can find the bending angle of a charged static EMD black hole according to
\begin{equation}
\delta =2\int_{0}^{u_0}\bigg|\frac{d\phi}{du}\bigg|du-\pi= 2\int_{0}^{u_0}\frac{1}{\sqrt{\frac{(1-r_-u)^{\frac{4\alpha^2}{1+\alpha^2}}}{b^2}-(1-r_+u)(1-r_-u)u^2}}du-\pi,
\label{32}
\end{equation}
where $u_0=\frac{1}{r_0}$ is the inverse of the closest distance. We have numerically plotted the behaviour of the bending angle as a function of $u_0$ in Figure 13. From the left panel of the figure, it is clear that by increasing the dilaton coupling, the bending angle also increases, so that a GMGHS black hole with $\alpha=1$ has the largest value of the deflection angle whereas the Reissner-Nordstrom black hole with $\alpha=0$  has the smallest value (see also Ref. \cite{new93}).  In Table 2  we see that by increasing $\alpha$  the event horizon, photon and shadow radii increase too. This shows that increasing $\alpha$ enhances the gravitational field and thus the deflection angle also increases with increasing the dilaton parameter. The dependence of the bending angle on the charge parameter is presented in the right panel of Figure 13, showing that for a fixed value of $\alpha$, when the value of the charge parameter increases, the bending angle decreases. The corresponding results for the Reissner-Nordstrom  and  Schwarzschild black holes in GR have also been plotted. Moreover, as one expects, in the limit of $u_0\rightarrow0$ the bending angle is zero while, for larger values of $u_0$ the deflection angle increases.

\begin{figure}[H]
\centering
\includegraphics[width=2.75in]{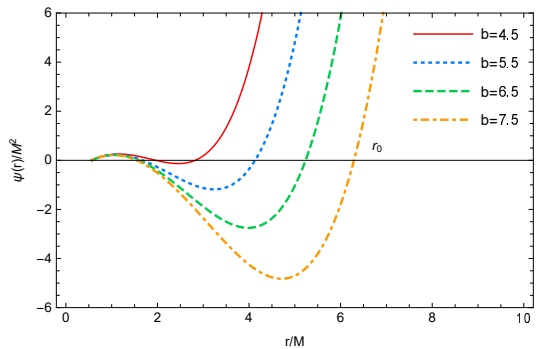}
\caption{\footnotesize The behaviour of $\Psi(r)$ as a function of $r$ for different values of the impact parameter. We set $\alpha=0.1$ and  $v=0.9$.}
\label{closest-distance}
\end{figure}

\begin{figure}[H]
\centering
\includegraphics[width=3.0in]{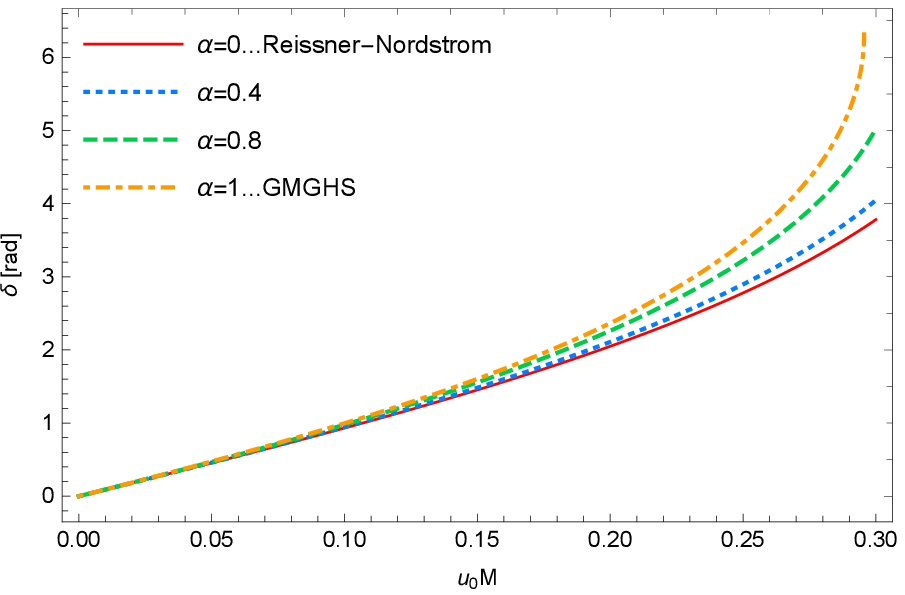}
\includegraphics[width=3.0in]{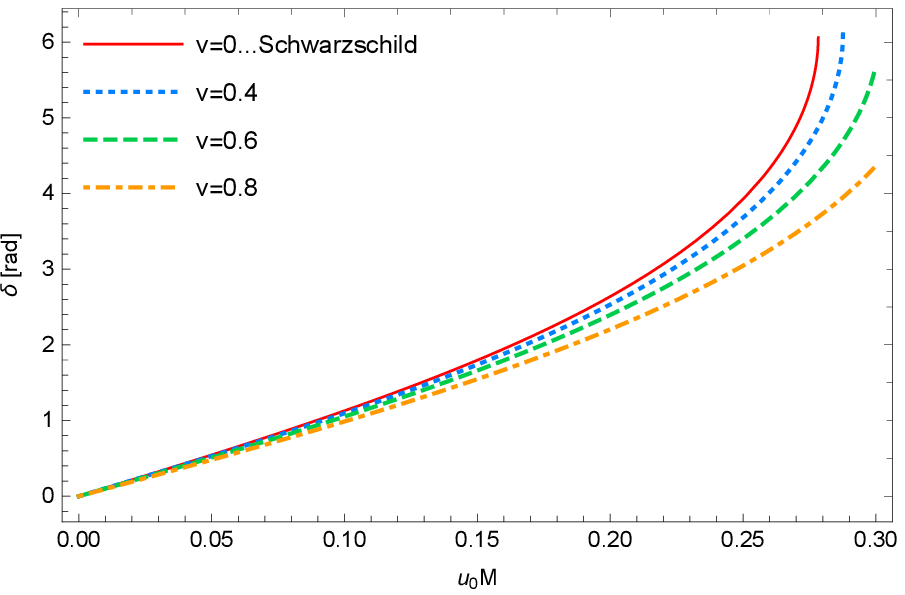}
\caption{\footnotesize The bending angle of charged static dilaton black holes as a function of inverse of the closest distance, $u_0=\frac{1}{r_0}$, for different values of $\alpha$ with $v=0.9$ (left panel), and for different values of $v$ with $\alpha=0.1$ (right panel).}
\label{deflection-angle}
\end{figure}

\section{Conclusions}
In this paper, we have studied the null geodesics structure of charged static dilaton black holes in EMD gravity for arbitrary values of the dilaton coupling constant $\alpha$. A detailed analysis of the null geodesics around a GMGHS black hole with a specific value of $\alpha=1$ had been done by Fernando in \cite{GMGHS1}. However, in the present work, we have considered EMD black holes with arbitrary values of $\alpha$ and investigated the effects of both the dilaton coupling and charge parameter $v$ on the null geodesics around them. For non-radial null geodesics we numerically obtained the effective potential, effective force on the photons,  radius of the photon sphere and its impact parameter. We found that for a given value of the charge parameter,  increasing the dilaton coupling $\alpha$ causes the photon radius  to increase. Therefore, in the presence of dilaton hair, a black hole in EMD gravity has a larger photon radius  compared to a Reissner-Nordstrom black hole (the scalar free solution) in GR. Also, we showed that for a fixed value of $\alpha$, by decreasing the charge parameter, the photon radius increases so that a Schwarzschild black hole with $v=0$ has the largest value of the photon radius.

It was also shown that depending on the photon's energy level, there can be different kinds of motion. In this regard, by considering the incoming light rays from infinity we discussed the possible motions of  photons around EMD black holes. We showed that photons with a critical value of impact parameter, $b_c$, move along the unstable circular orbits, while,  photons with $b<b_c$ eventually fall into the black hole. In a different scenario, we considered the unbounded orbits for light rays with $b>b_c$. The null trajectories of these three cases were also numerically plotted. In addition, we have studied the shadow cast by charged static dilaton black holes. We found that in the presence of the dilaton coupling the size of the black hole shadow increases, while the shadow radius would decrease when we increase the charge parameter. Indeed, it was shown that the shadow size has a strong dependency on the charge parameter $v$. From Table 1 we see that  for $\alpha=0.2$ and a varying $v$ from 0 to 0.9, the shadow size decreases by about 0.871. However, the dependency on the $\alpha$ parameter is very weak, as can be seen from Table 2 where for a fixed value of $v=0.6$ and a varying $\alpha$ from 0 to 2, the size of shadow increases only by about 0.037. It would therefore be difficult to measure the  parameter $\alpha$ by observational measurements of the shadow. Finally, we studied light bending in the space-time of EMD black holes and investigated the effects of the charge parameter and dilaton coupling on the bending angle. The results were compared to dilatonic GMGHS black holes with $\alpha=1$, Reissner-Nordstrom black hole with $\alpha=0$ and Schwarzschild black hole in GR.

\end{document}